# Knowledge representation and scalable abstract reasoning for simulated democracy in Unity

ELEFTHERIA KATSIRI[1,*], ALEXANDROS GAZIS[1], ANGELOS PROTOPAPAS[2]
[1]Department of Electrical and Computer Engineering, School of Engineering,
Democritus University of Thrace,
67100 Xanthi,
GREECE
[2]Department of Civil Engineering, School of Engineering,
Democritus University of Thrace,
67100 Xanthi,
GREECE

*Abstract:* We present a novel form of scalable knowledge representation about agents in a simulated democracy, e-polis, where real users respond to social challenges associated with democratic institutions, structured as Smart Spatial Types, a new type of Smart Building that changes architectural form according to the philosophical doctrine of a visitor. At the end of the game players vote on the Smart City that results from their collective choices.
Our approach uses deductive systems in an unusual way: by integrating a model of democracy with a model of a Smart City we are able to prove quality aspects of the simulated democracy in different urban and social settings, while adding ease and flexibility to the development. Second, we can infer and reason with abstract knowledge, which is a limitation of the Unity platform; third, our system enables real-time decision-making and adaptation of the game flow based on the player's abstract state, paving the road to explainability.
Scalability is achieved by maintaining a dual-layer knowledge representation mechanism for reasoning about the simulated democracy that functions in a similar way to a two-level cache. The lower layer knows about the current state of the game by continually processing a high rate of events produced by the in-built physics engine of the Unity platform, e.g., it knows of the position of a player in space, in terms of his coordinates x,y,z as well as their choices for each challenge. The higher layer knows of easily-retrievable, user-defined abstract knowledge about current and historical states, e.g., it knows of the political doctrine of a Smart Spatial Type, a player's philosophical doctrine, and the collective philosophical doctrine of a community players with respect to current social issues.

## 1 Introduction

This work reflects on the design and implementation of an in-engine multi-agent simulation in Unity that interacts with real players, e-polis. e-polis is a serious game and a methodological tool that integrates a novel knowledge representation and reasoning framework in order to investigate a) the way young people think and act in today's social and political arena and b) the way they imagine the ideal community (polis) [1]. In order to draw as representative conclusions as possible, such a serious game needs to cater for diversity, equity and inclusion (DEI), be able to scale out to large segments of the population and be as engaging as possible, i.e., it requires an interdisciplinary approach. Furthermore, as society evolves it needs to be able to adapt to current affairs. However developing a serious game that acts as a research tool requires high level by programming abstractions that are not supported by modern game development environments, such as Unity, which is a distributed system where each game object lifecycle is programmable via C# scripts that model the object's monobehaviour

[1]This serious game and methodological tool is part of a wider research project called "e-polis of the future". The research programme, its theoretical (sociological, epistemological and architectural) foundation, the architectural drawings and the game's implementation with data extraction were carried out by a research team of the National and Kapodistrian University of Athens, consisting of: Orestis Didimiotis, Sara Gogou, Orestis Konstantas, Gerasimos Kouzelis and Despina Paraskeva-Veloudogianni. The full description of the project has been published in the volume by the same authors, *Representations of Democracy: Responses from a Youth Survey*, Athens 2025, Nissos-EMEA Publications.

(Appendix A.4). As a result, the programming environment has very limited support for global variables and global classes, which, together with the inherent lack of persistent storage, is not adequate for reasoning with abstract global and historical state. For example, Unity on its own cannot be programmed in order to evaluate e.g., the consistency of the players' philosophical belief over a series of actions (answering the political dilemmas) or infer the prevailing philosophical doctrine of all players with respect to religion.

Furthermore, the following issues further hinder the development of a serious game:

### Content management

- Creating parametrical Smart City layouts out of custom building blocks of different sizes and types. For example, in the scope of this work, for each combination of urban block area, road width, density and height distribution of the buildings, layout components (neighborhoods of initially empty urban blocks) were created manually, replicated and puzzled together in predefined, calculated positions on the main scene to form a Smart City. Second, for each social institution of interest, five different architectural versions were developed in Rhino, exported as generic objects, imported and extracted in Unity, edited, cleaned, re-packaged as Unity components per Spatial Type and philosophical doctrine.
- Creating a large number of up-to-date social and political challenges from custom textual and multimedia content. For example, in the scope of this work, a challenge template was created manually as a UI object containing 5 buttons, replicated 30 times, populated with images and text) and placed in predefined, calculated positions on the challenge scene for the user to interact with. Other types of multimedia such as video and animation were desirable. As social and political issues evolve, new challenges need to be created, others updated or removed following the above process.
- Creating scripts (in C#) for each game object, i.e., layout components, smart spatial types, UI buttons and attaching them to each object.

### Game logic

- Creating algorithms for calculating abstract game state predicates that reflect research goals from low level events produced by the Unity Physics Engine. For example, the game needs to be able to infer when one or all players have responded to all challenges, which of the challenges they were most interested it, what is a player's philosophical doctrine based on the choices they provide to challenges, how happy are the players from the resulting Smart City.
- Creating algorithms for predicting abstract state predicates for new or current players based on the collected data.

### Scalability and expressiveness

- Managing big data generated by multi-player versions of serious game. A very large number of events are generated in real-time from tracking user movements and their interactions with game objects that are required by the game logic that need to be processed in real-time and in order to perform actions and produce the research outcomes.
- Exporting the game for accessibility. The game needs to be made publicly available in a repository, with all security patches applied, from where it can be downloaded easily and installed as an app.
- Creating an analytical tool for processing user choices and generating research outcomes. The results should be properly annotated and collected in a cloud service where they can be analysed using advanced data science techniques.

Tasks above are pedantic, time consuming and prone to error and raise the need for a model-based approach that will improve the reliability of the serious game, while decreasing the burden of the development. The approach implemented in this work, decouples the abstract reasoning from the game mechanics, allowing different implementations of the same game in different domains, such as cities with different density and island communities, supports both current and historic predicates that can be analysed to determine trends and other statistics, and enables real-time decision-making based on the player's abstract state. In the rest of this paper, the terms *challenge*, *dilemma* and *pentalemma* are used as synonyms.

## 2 Related work

Although few approaches using smart cities in serious games exist in the literature, their aim has been to investigate Smart Cities, their aspects and their problems, [1].

With the advancement of Digitalisation, participation of citizens in governance started to increasingly become electronically enabled since the early 1990s, in hopes of facilitating a wider reach and a better inclusion of marginalized groups in governance and democratic practices, [2], [3], [4]. E-participation, [5], [6], can holistically be considered as technologically mediated interaction between the government and civil society, [7], towards the betterment of their community, [8].

Digital games are complex socio-technological artifacts that are hard to define, [9]. They often represent artifacts created for entertainment purposes, where users are faced with artificial challenges, rules for solving those challenges and outcomes from engaging with the challenges, [10]. Playing games has been associated with several cognitive, emotional, motivational and social benefits, [11]. Hence, games, in a form or another, have been utilized in policymaking and related activities for decades so as to harness these benefits of games in non-game contexts, [12], [13], [14], [15]. Gamification is an umbrella term for the use of video game elements (rather than full-fledged games) to improve user experience and user engagement in non-game services, [16]. The growth of gamification accompanied the expansion of smart phones and mobile apps, which allow access to various services via a digital interface, [17]. Gamification has proliferated as a tool for democratic participation – notable examples include participatory budgeting in Argentina and Canada, [18], urban planning in Finland, [19], and civic discussion of social issues in Nigeria, [20]. Scholars of participatory democracy argue that controlling contingencies through game mechanics reduces the high levels of user boredom, frustration, and disengagement while producing more empowered democratic processes, [18], [21]. Decide Madrid and vTaiwan integrated game elements into their interface design to address also online 'trolling' (irritating other players), and polarisation caused by social media, [22]. The policy gaming field, for one field bridging games and governance, looks into, how simulation games can assist in policy planning and better organizational decision-making, [23].

Game-based democracy, utilises certain design rules to lead and provoke users to become active and engaged, and to feel enjoyment and fun. 'Game-based elements', that are central to the process of gamification, [17], [24], generally include two characteristics: rule-based competition and incremental rewards building toward a specific goal, [16], [18], [24]. During a rule-based competition, users are rewarded with points, badges or other incentives for completing designated tasks towards a clearly defined goal, [18], [25].

CityCare utilizes a point system, on top of a problem reporting system, to encourage citizens to actively communicate any problems they encounter in their city to administrators, [26], TAB Sharing, [27], is such a web-based and mobile platform where citizens submit proposals and solutions to obtain points and enter different kinds of leaderboards. Additional points are received for each vote received by other users and from positive feedback from the administration. Medals enable citizens to change their level while "distinguished" citizens are rewarded publicly every year.

Other gamification strategies include creation, management and use of flashcards, games, quizz games, maps, resources, quests, challenges, simulations, presentations, infographics, posters, catalogs, images, playsheets, etc., [28].

Gamification is often viewed as a form of control of contingencies, construed in the context of democracy as challenges, [18], [21]. A good game design produces an 'artificial conflict' in which users (either individuals or groups) compete against each other based on predetermined rules, [18].

The elements of feedback, progress tracking, and an ever-present sense of achievement present in gamified environments contribute towards immersive and enjoyable educational experiences, [29].

By design, in vTaiwan's digital interface creation of 'artificial conflicts', is produced through the purposeful visual display of (different) Opinion Groups on the user interface. The interface visualises how many users support and oppose each Opinion Group, positioning users in conflicting positions. The design is supposed to 'nudge' participants to explore and reflect on the opposite group's opinions (for each issue under discussion). These methods are different from a traditional online forum, which only rewards the most 'liked' opinion, [22]. Decide Madrid's design interface establishes a set of complex goal-oriented rules and emphasises 'competition' amongst participants over generating a consensus from e-petitions to a participatory budget process funded by public revenue, [18].

Geographers refuse the reductionist framing of videogames as a space where relationships between users and videogame technologies are predetermined by game designs. Rather, videogames are understood to be open-ended environments/spaces of emergence, unfolding, and becoming as they are experienced by users, [30], [31], [32]. For [32], designers of videogames have to manage and anticipate (but not completely control) Contingencies, i.e., unpredictable interplays between users and the gamified interface are managed but not completely controlled as they generate and encourage creative responses from players. However, a gaming environment that is too open might 'fail to captivate and capture an audience or community of users at all', [32], [33].

Gamification can be implemented through heavily relying on the introduction of design features unique to games (known as game elements) to existing systems or service, [16], [34].

Gamification implementations can also extend beyond the digital, employing physical props/hardware such as game controllers, such as in, for example, an application intend to assist law enforcement officers in the regulation of drones, [35]. Gamification can also be implemented through stories and role-play that immerse players in imagined realities, such as one where the players are mapping a city to facilitate the development of accessible mobility maps while fighting off zombies, [36]. Due to these differences in its implementations, outcomes from gamification can vary and hugely depend on the implemented design and the use context, [37].

Studies of gamification offer one view of digital democracy, but they tend to paint a rather linear and techno-solutionist picture. They ‘do not consider the impacts of various failures, glitches and contestations in user-interface relationships known as friction. Emotional predicaments such as frustration, confusion, or disengagement are an unavoidable part of gamified participation, and the digital interface sometimes loses control of participants’ emotions and behaviour.

Teachers and game designers can fail to establish differences between Serious Games, Game-Based Learning or Gamification. As [38], indicates, Serious Games start from a real problem and include it in a game to make it more fun and easier to understand. In Game-Based Learning learners play games (digital or non-digital) in order to learn contents, [39]. Examples of resources for Game-Based Learning are: World Peace Game Foundation, World of Warcraft in School, Minecraft Edu or Portal 2. Gamification is not a game, but it applies the techniques and elements of games in order to motivate, encourage changes in attitudes, improve processes, etc. It is not based only on rewards, points or rankings, but implies a pre-action analysis process, [16].

The seamless integration of technologies, ethical considerations, accessibility, education for technological literacy, interoperability, user trust, environmental sustainability, and regulatory frameworks are becoming significant. These challenges present opportunities for the future to enrich human experiences while addressing societal needs.

In parallel, though, a more disruptive interaction paradigm is taking place, rooted in immersive technologies such as augmented, virtual, and mixed reality (AR, VR, and MR). Apart from gaming special-purpose applications (e.g., in museums, science centers, 4D cinemas, etc.), these interaction paradigms exist in a newborn universe.

The advent of immersive technologies, [40], [41], AR, [42], VR, [43], and MR, [44], made possible the transition from traditional screens to the digital three-dimensional space, supporting both hybrid and pure digital training experiences. XR applications in vocational education and training exemplify the capacity of these technologies to revolutionize experiential learning, [45], [46], [47].

In such applications, human motion capture, [48], [49], and 3D digitization, [50], [51], [52], enable users to interact with digital environments in ways that mirror real-world movements and real-world environments, [53], [54], [55]. Such realistic simulations blur the boundaries between physical and digital realities, fostering a deeper level of engagement and personalization in digital experiences. Immersive technologies can also enable new forms of social interaction, more realistic and engaging virtual meetings and conferencing, dynamic virtual versions of physical places or objects, known as “digital twins, [56], that can model complex systems, enable dynamic, real-time design and collaboration. Immersive games are video games that transport the player into an alternative world where techniques are used to make them feel more like the character they’re playing. As the compute power of smartphones increases and 5G becomes prevalent, mobile latency and buffering issues are diminishing, [57], allowing game studios to create seamless gaming experiences for users, even as they transition between consoles, laptops, and mobile devices. Arm is at the center of immersive gaming experiences from the cloud to the edge and endpoint while companies such as Hatch, Google, Epic and Unity optimize mobile gaming performance on Arm-based SoCs.

Serious games can be considered as the domain where user-centric principles and the innovation described in AR, VR, and MR join forces to re-purpose gaming and address educational, training, and societal challenges. Serious games offer purposeful play that transcends entertainment, becoming powerful tools for knowledge and skill acquisition and thus addressing broader societal issues. Serious games employ XR technologies and game design principles to address educational, training, and societal challenges, [58]. These games are based on a paradigm shift from entertainment-focused gaming to skill development, knowledge acquisition, and societal awareness, [59]. At the same time, the integration of artificial intelligence (AI) into UIs, [60], [61], empowers systems to understand user intent; process information efficiently; and offer personalized, anticipatory digital experiences, [62].

Ontologies, linked data, and semantic graphs provide a rich framework for expressing relationships between concepts, [63], allowing systems to infer and connect pieces of data, creating a web of contextual relevance. Semantic knowledge representation lays the groundwork for a more intuitive and context-aware service provision, [64]. NLP algorithms, powered by semantic models, enable systems to comprehend user queries in a more human-like manner, [65]. Beyond representation, the presentation of information is an important part of user experience. Semantic technologies should influence how information is visually and contextually communicated to users, [66], ensuring that users receive information in a format that aligns with their cognitive processes and preferences and has the potential to provide a more coherent and holistic user experience, reducing information overload and enhancing overall satisfaction.

# 3 Background

Human–computer interaction (HCI) and information and communication technologies (ICT) are experiencing a transition due to the emergence of innovative design paradigms that are affecting the way we interact with digital information. Immersive technologies, such as augmented reality (AR), [67], [68], virtual reality (VR), [43], and mixed reality (MR), [69], are considered the building blocks of modern digital experiences. Knowledge representation in a machine-interpretable format supports reasoning on knowledge sources to shape intelligent, user-centric information systems. From a human-centered perspective, these systems have the potential to dynamically adapt to the diverse and evolving needs of users. Simultaneously, the integration of artificial intelligence (AI) into ICT creates the foundation for new interaction technologies, information processing, and information visualization. On the other hand cities of tomorrow need to adopt a holistic model of sustainable urban development. The implementation of a smart city is strongly related to the process of social (geo)participation—according to PAS 180: 2014 (Publicly Available Specification under the British Standards Institution), this process means an "effective integration of physical, digital and human systems in the built environment to deliver a sustainable, prosperous and inclusive future for its citizens." Also, the new urban agenda stresses the need to empower all individuals and communities and to promote and broaden inclusive platforms that allow full and meaningful participation in the decision-making and planning process, [70], [71]. In recent decades, the crisis of participatory democracy has been particularly severe in urban centers and areas subject to urbanization. Its outcome is the weakening of the sense that the residents have a real impact on co-creating the vision for a city's development, revitalization of the neglected districts, or spatial order. When interpreting the term "the right to the city", [72], emphasizes that it is not only the right to access its resources, but also the right to decide jointly on the direction of the city's development. A smart and sustainable city engages all of its residents in the most critical decision-making processes, making the process of creating spatial order more social, and encouraging the development of participatory and deliberative democracy. This work discusses a framework and a novel methodological approach for game-based democracy in a Smart City State context that abides by the above principles. The open problem to address is to examine the way in which youth, as a critical agent of social change, conceptualise democracy today, both as a concept and as a set of political practices, thus anticipating the future society. To this end, it examines in combination a) the way young people think and act in today's social and political arena and b) the way they imagine the ideal community (polis). More specifically, the purpose of this article is to develop the concept of smart cities' residents as active *urban sensors* represented by player avatars. Consequently, both city residents and spatial types are considered elements of a specific geospatial immersive system. The impact the "human agents" have on spatial objects in the long-term influences the the spatial planning of the city, while mutual interactions of the residents, between the residents, combined with a visual feedback loop over the resulting "ideal" city state, influences further their democratic viewpoint. Therefore, the goal of the authors is to model complex social interactions between the urban residents and the smart City-State and to map the level of young people's attitudes towards established institutions, practices and perceptions, while freeing them to envision and create the ideal society, the expected social relations and the governance practices within it. The conclusions of the research are based on the analysis of the "ideal city-states" to be developed in the game in combination with the findings of the mapping. Based on this approach, the project further explores which aspects of youth's socio-political daily life prefigure trends of transformation - and in which direction - how young people tend to resolve political crises, how they evaluate the concepts of inclusion and solidarity, how reality influences fantasy and vice versa, and finally how the concept of utopia is currently conceptualised. To this extent, this article introduces a novel game element: a *Smart Spatial Type* that changes its architectural shape according to the political doctrine of the person that interacts with it, while this user strolls in the Smart City; Physical world *Spatial types*, an urban planning term, is where human activities, the equitable access of citizens to services and infrastructure and the prudent management of natural resources and cultural resources are placed; these are supported by legislative provisions and the development of a comprehensive set of policies. The Smart Spatial Type therefore, acts as a *digital interface* to physical building blocks, such as Government buildings, Parks, Schools, Theatres, Temples, Markets. A *socio-political dilemma*, linked to the Smart Spatial type, i.e., a depiction of a "social scenario" with five choices each representing a different philosophical doctrine. The *Smart Spatial Type* is transformed to a different architectural form, as a result of the player's position in the dilemma. User positions are recorded in a cloud database for further analysis. At the end of the game the users can watch the city they have created from above and contemplate on how their answers contributed on reforming their city. A *Smart City-State* is a smart city that is structured as a collection of *Smart Spatial Types*. It represents a city that can be considered a smart one when it, in parallel, invests in technology and human capital to actively promote sustainable economic development and high quality of life through civic participation. The e-polis concept is analysed in detail in the next section.

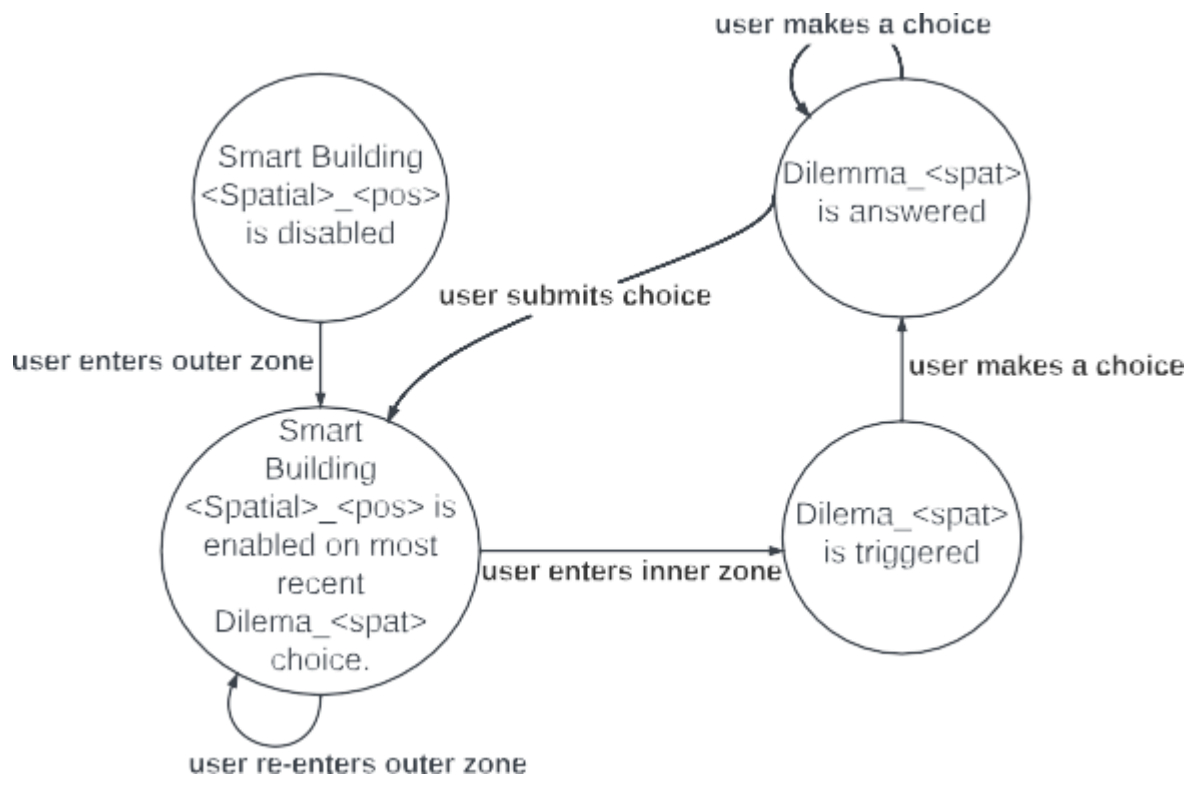

Figure 1: Smart Spatial Type concept

## 4 Concept

In this work we introduce the concept of a *SmartSpatialType* that has a state and two *active* zones and uses the following modus operandi: The outer zone includes approximately the whole building block while the inner zone contains approximately the building infrastructure. Initially, the state is equal to disabled, i.e., the building block appears to be empty. The first time the user enters the outer zone, a default instance of the *SmartSpatialType* is enabled, i.e., appears suddenly. If the users continues their approach, once they have entered the inner active zone, they get transported to a linked location where they are faced with a social five-choice dilemma. In order to continue the game they have to select at least one of the available choices and then press the Continue button. The choice, which is logged both in the cloud database and persistently within the game Player Preferences, causes the *SmartSpatialType* to change to a predefined shape that is linked with the *UserChoice*. Last, the user is transported to the original location to inspect the result and continue the game. This sequence of user-building interactions is described schematically in Figure 1.

## 5 Novelty

Our modelling approach in this work leverages a dual-layer knowledge representation architecture for *Sentient computing*, developed in previous work, [73]. Sentient computing, [74], is the proposition that applications can be made more responsive and useful by observing and reacting to the physical world. The dual-layer knowledge representation architecture makes it possible for the user to perform easily complex computations involving spatial and temporal notions of a dynamic changing environment. This work builds on the previous system, where Sentient Computing predicates represent *location-awareness*, extending it with first order logic predicates that represent knowledge about the simulated democracy, such as when one or all players have responded to all challenges, which of the challenges they were most interested it, what is a player's philosophical doctrine based on the choices they provide to challenges, how happy are the players from the resulting Smart City.

More specifically, the *Deductive Abstract Layer* in this work has abstract knowledge about the player's preferred *political position* in specific socio-political dilemmas. It also has knowledge on *user intent*, i.e., whether the player *passes by* a Smart Spatial Type, or *visits* it. Using AI on historical knowledge predicates, the overall political position of a player can be inferred, including similarities among players creating *opinion groups* such as the ones in the vTaiwan. Furthermore, the *popularity* of a Smart Spatial Type can be inferred, both for a specific user, and for all users.

## 6 System architecture

In this work, Smart Spatial Types were modeled as *urban sensors* (Figure 2). Smart Spatial Types periodically generate events, that are asserted in the Sensor Abstract layer (SAL) as low-level predicate instances that represent low-level knowledge of

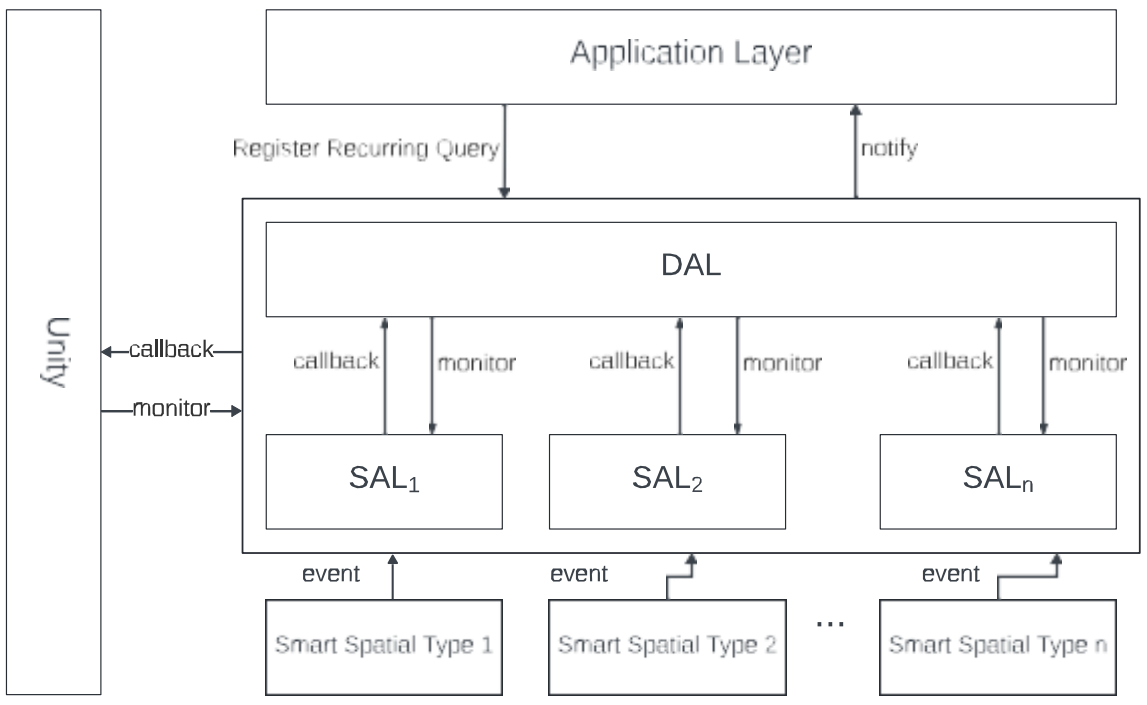

Figure 2: System reference architecture

- player collision with one of the two boundaries of the Smart Spatial type either at the outer boundary (the outdoor area) or the inner boundary (the indoor area) where the challenge is triggered, at the current instant
- player interaction with a UI button that provides an answer to a given challenge, associated with the Smart Spatial Type that the player has entered.
- player voting on the current Smart City state at the end of the session

A new predicate instance is created every-time a player enters a Smart Spatial Type, triggers a challenge, provides a choice or a vote. For each Smart Spatial Type, SAL knows of all recent collisions of a player with each boundary as well as all given choices to a challenge. Every time an assert action is performed the SAL layer matches consecutive predicate instances against patterns that indicate abnormal behaviour. For example, random random movement, long periods of inactiveness, repeated re-entering of the Spatial Type, repeated choices for the same challenge, that may indicate a special interest on the player's side for the specific building or associated challenge or give other insights.

With respect to the Deductive Abstract Layer (DAL), users register recurring query specifications that represent high-level knowledge about the outcomes of the players interactions with the Smart Spatial Types. Each query is structured as a Rete network described in Section 7. It monitors continually low-level predicates residing in SAL causing new instances of the DAL predicates to be asserted, using a callback mechanism. DAL predicates are available for further querying by applications and they can be composed into recurring or single queries to meet application requirements.

For example, for a given Smart Spatial Type, the DAL layer knows the philosophical doctrine of each choice provided by each player for each challenge, the most popular philosophical doctrine for each challenge and for all the instances of the Smart Spatial Type. Accross Smart Spatial types, for a given player, the DAL layer knows of the most frequent (predominant) philosophical position for this player, whether a choice diverges from it and it can predict upcoming choices, which can give useful insights in the research. Last, for all players, it knows of the prevailing philosophical doctrine of their choices as well as the prevailing opinion on the state of the Smart City an any time.

With respect to the *Unity platform* 2 it complements the knowledge base to which it is linked, by offering two services:

- It generates low level events using the Unity event system and the UI event system and asserts them into the SAL layer.
- It adapts the flow of the game to the context of the player. By listening for callbacks from the knowledge base and responding with an appropriate action, it is able for example, to transport the player to the top of the main scene to see the city from above or transport them to a location i in the game where they can vote on an particular challenge for example if additional information is required.

## 7 Technical description

The inference is based on the *Rete algorithm*, a very efficient algorithm for matching facts against patterns in rules. It is implemented by building a network of nodes. It is designed in such a way that it saves the state of the matching process from cycle to cycle and re-computes changes only for the modified facts. The state of the matching process is updated only as facts are added and removed and the fewer they are the faster the matching will be. Each rule has an inference cycle consisting of three phases: match, select and execute. In the matching phase, the conditions of the rule are matched against the facts to determine which rules are to be executed. The rules whose conditions are met are stored in a list called agenda for firing. Form the list of rules, one of the rules is selected to execute or fire, according to a selection strategy, e.g., based on priority, recency of usage, specificity of the rule or other criteria. The rule selected from the list is executed by carrying out the actions in the right hand side of the rule. The action may be an assertion, executing a user-defined or built-in function. The main advantage of Rete is speed as it takes the advantage of structural similarity in rules. On the other hand, its main drawback is that it is memory intensive as saving pattern matches and partial matches requires considerable amounts of memory.

A *Rete network* is a direct acyclic graph that consists of nodes representing patterns in the conditions of the rules. The nodes behave like filters, testing the incoming tokens and sending on only those that pass the test. The Rete network consists of two parts: alpha network, consisting of nodes known as alpha nodes and beta network, consisting of beta nodes. Alpha nodes have one input and evaluate the conditions of the rule where beta nodes have two-inputs and define inter-element (fact) conditions. Each alpha node is associated with an alpha memory that stores the matching facts. Alpha nodes are then joined by beta nodes that only have two inputs thus supporting partial matching. Each beta node has a memory to store joined patterns. Assertion of a fact creates a token that initially enters the root node. The network then splits a branch for each token type. Each kind node gets a copy of the token and preforms a SELECT operation to select tokens of its kind. It then passes a copy of the token to the alpha nodes. On receiving the token the alpha nodes do a PROJECT operation and and extract components from the token that match the variables of the pattern, thus evaluating the pattern. Beta nodes then determine the possible cross-products of the rule. Finally, actions in the rule will be executed.

A custom embedded *middleware component* running on the player device where the serious game is installed, links the system components together using a text-based interface which involves exchanging text via a shared file space, PlayerPreferences. Unity writes the events on the shared space making a callback to the knowledge base that converts them to the correct format in order to assert them in the knowledge-base. Vice versa, the knowledge base exports predicate instances in the form of C# boolean variables that are picked up by Unity scripts causing the desirable action.

The *clipspy* bindings, a “pythonic” thin layer built on top of the CLIPS native C APIs was used to analyse some of the collected data providing tools such as plots and clustering. Furthermore, *Real-time game data* is collected in the Firebase noSQL cloud database. The data can be accessed through the Firebase console, at the FirebaseEPolis project. To interface Firebase with the game, the Firebase Unity SDK was integrated into the e-polis bundle. The SDK supports both a desktop workflow and mobile workflows in Android and IOS.

# 8 Knowledge Representation

The basic element of the model is the *SmartSpatialType*, that comprises of:

- a *SpatialType*.
- a 2-order CompositeLocation.
- a set of $n$ (Politicised)SpatialTypes, each are located at the CompositeLocation and that only one instance is enabled at each time.
- An AtomicLocation that is not associated with the Composite location.
- A Dilemma that is located at the above AtomicLocation.

An *AtomicLocation* is a polyhedral region that has a centre and a polyhedron. A *SphericalAtomicLocation* is a version that has a spherical region instead of a polyhedron. A *CompositeLocation* is a set of two or more nested AtomicLocations that have the same centre and different polyhedrons, such that one contains the next, and so on. A *2-order CompositeLocation* contains 2 nested polyhedrons. A *SpatialType* represents a type of public space with infrastructure and a function, e.g., a factory spatial type. A SpatialType instance, e.g., the Cadbury’s factory in Birmingham, has a state that can be either enabled or disabled. A *(* Politicised)SpatialType is a spatial type that is influenced by a given philosophical dogma. For example a Monoblock type of appartment building is a politicised specialisation of the *SpatialType* AppartmentBuilding associated with Communism. A *PolyhedralSpatialType* instance is a 3D object were the x-z sides are significantly larger than any x-y or z-y side. That is the height is significantly smaller than the width and the length making it a flat 3D object. It is used to model public locations such as roads that do not belong to any *SpatialType*. A *PoliticalType* represents a philosophical doctrine such as Conservatism. A *Dilemma* is an *AtomicLocation* where a *UserChoice* predicate instance can be asserted. A *UserChoice* represents the user’s preferred philosophical position, e.g., CL, on a social issue at hand, represented by the dilemma, e.g., a forgiving a theft in the marketplace as inevitable due to social injustice. *UserChoice* is also associated with a SpatialType instance. Given a UserChoice as above, the associated SpatialType instance that corresponds to the User’s preferred theory is enabled and the previous instance is disabled. This has the effect that the user can see the impact of their choices on the city.

A *PoliticisedSpatialType* is essentially a specialisation of a*SpatialType* with an extra attribute that corresponds to a *PoliticalType*. The size of a *SmartSpatialType* is equivalent to the $x$-$z$ area of the smallest polyhedral of the *CompositeLocation* it contains. There are three sizes of SmartSpatialTypes:

Single, Double, Quadraple. A *RoadBlock* is a *SpatialType* that comprises of:

- a 2-order CompositeLocation.
- an instance of a PolyhedralSpatialType.

The *RoadBlock* can be further gamified , leveraging the *CompositeLocation* element, to by enable it only as user approaches it, creating the illusion of appearance.

An e-polis is a *SmartCity* that is structured as a collection of *SmartSpatialTypes*, *RoadBlocks* and a layout. A layout is a map of the centres of the *CompositeLocations* of the *SmartSpatialTypes* and the *Dilemmas*. The dimensions (size) of the Single *SmartSpatialType* can be parameterised initially to create different building density and shape of the Smart City.

# 9 Formal specification

The key logical predicates are shown in this Section.

## 9.1 Sensor Abstract Layer

```
(L_AtomicLocation (uid ?user_id)
(x ?x)(y ?y)(z ?z))
(L_AtomicLocation (rid ?region_id)
(sid ?sid)(polyhedron))
(L_NestedLocation
(nrid ?nested_region_id)
(prid ?parent_rid)(sid ?sid)
(polyhedron))
(L_SpatialType (sid ?spatial_id))
(type ?type))
(L_PoliticisedSpatialType (sid ?sid)
(sid ?spatial_id)(pid ?dogma_id))
```

## 9.2 Deductive Abstract Layer

```
(H_ User At Nested Location (uid ?uid)
(rid ?rid)(start_time ?time_value))
(H_User At Smart Spatial Type (uid ?uid)
(sid ?sid)(start_time ?time_value))
(H_UserInDilema (uid ?uid)
(did ? dilemma_id )
(start_time?time_value))
(H_UserAnsweredDilemma (uid ?uid)
(did ?did)(choice ?choice_id)
(start_time?time_value))
(H_SmartSpatialTypeTransformed
(sid ?sid)(pid ?dogma_id
(start_time?time_value))
(L_Dilemma (did ?dilemma_id )
(sid ?spatial_id))
```

## 9.3 Historical predicates

```
(H_UserInLocationHist(uid ?uid)
(siid?siid)(sid ?sid)
(start_time ?time_value)
(end_time ?time_value))
(H_ User In DilemaHist (uid?uid)
(did ?did)(start_time ?time_value)
(end_time ?time_value))
( H_ User Choice Hist
(uid?uid)(did?did)
(choice?choice_value)
(start_time?time_value)
(L_ User Choice (cid?choice_id)
(uid ?user_id)(did ?dilemma_id)
(sid?spatial_id))
(L_PoliticalDogma(pid ?dogma_id)
( name ?name ) )
(L_PoliticisedUserChoice
(cid ?choice_id)
(pid ?dogma_id))
```

# 10 Example

The prevailing philosophical doctrine for a challenge is the most frequent ”political flavor” of the players’ choices for that challenge. The following CLIPS code illustrates how the prevailing philosophical doctrine for the challenge it can be computed by the DAL layer. Assuming that the following predicate instances are asserted in DAL as a result of the choices of different users for a given challenge (did 1), the most prevailing philosophical doctrine is democratic realism that is represented by the two instances of choice 1.

```
(H_UserAnsweredDilemma (uid 1)(did 1)
(choice 1)(start −time 1000)
(H_UserAnsweredDilemma (uid 2)(did 1)
(choice 1)(start −time 1020)
(H_UserAnsweredDilemma (uid 3)(did 1)
(choice 3)(start −time 1080)
(H_UserAnsweredDilemma (uid 4)(did 1)
(choice 2)(start −time 1100)
(H_UserAnsweredDilemma (uid 5)(did 1)
(choice 4)(start −time 1200)
(L_PoliticisedUserChoice ((choice 1)
(pid 1))
(L_PoliticalDogma(pid1)
(name democraticrealism)
(L_PoliticisedUserChoice (choice 2)
(pid 2)
(L_PoliticalDogma(pid2)
(name critical radicalism)
(L_PoliticisedUserChoice (choice 3)
(pid 3)
(L_PoliticalDogma(pid3)
```

```
( name apoliticism )
( L_PoliticisedUserChoice (( choice 4)
( pid 4 ))
( L_PoliticalDogma ( pid 4)
( name conservatism )
```

In order to calculate the most prevailing choice the following program first counts the instances of each type of choice and stores the count in instances of the H_SameChoiceDifferentPlayer predicate. Next, the instances are compared with respect to to the count slot in order to find the dogma that corresponds to the max count.

```
( defrule prevailing−choice
   (H_UserAnsweredDilemma ( uid ?uid1 )
   ( did ?did1 )( choice ?cid1 )
   ( start−time ?st1 ))
   ( H_UserAnsweredDilemma ( uid ?uid2 )
   ( did ?did2 )( choice ?cid2 )
   ( start−time ?st2 ))
   (= ?cid1 ?cid2 )
   => modify
   ( H_SameChoiceDifferentPlayer
   ( did ?did )( choice ?cid1 )
   (+ ?count1 ) ( start−time ?st1 ))

( defrule find−max−value
  ( H_SameChoiceDifferentPlayer
  ( did ?did )( choice ?cid1 )
  ( count?count1 )
  ( start −time  ?st1 ))
  ( not
  ( H_SameChoiceDifferentPlayer
  ( count ?count2 &:
  (> ?count2 ?count1 ))))
  =>
  ( assert ( H_PrevailingChoice
  ( did ?did )( choice ?cid )
  ( start−time ?st1 ))
  ( printout  t "Choice " ?cid
  "is␣the␣maximum" crlf ))
```

## 11 Prototype Implementation

This section discusses two different prototype implementations using the knowledge-base framework. The final version of the game is multi-player and includes 20 5-choice dilemmas which have been placed in the 20 empty (unbuilt) blocks, while in the remaining blocks that are not active, a selected PoliticisedSpatialTypeInstance is permanently visible as shown in Figure 3. The spatial types implemented are shown in Equation (1). To illustrate the gameplay with an example, let us assume that the player approaches enough the Factory 1 empty block: an apolitical version of a factory pops up as shown in Figure 4(a); as the player approaches more, a dilemma is triggered depicting a strike of factory workers with the options: "Without you, the Earth won't rotate!", "Strikes are so last year!", "Damn, there is no public transport; I will be late for work", "Amidst the economic crisis some people just keep asking for more" and "These trade unionists suck!"(Figure 5). Assuming the user selects the last option the spatial type is transformed to the one shown in Figure 4(b). After the player has answered all 20 dilemmas, they can see the city-state from above (Figure 6) and vote on the final outcome. The second prototype is in an island and contains three challenges, associated with an elementary school, a cultural centre and an orthodox church. Integrating the knowledge base improved the time needed to develop the business logic as the same CLIPS programs in the SAL and DAL layers were used and the Unity scripts associated with each game objects in the first prototype were replicated with small changes in the naming scheme and linked to the knowledge base. However, as the island terrain was significantly different to the urban one, a large number of the layout components were developed from scratch. One approach to address this issue trade-off between realistic depiction and usability is to use abstract layout components that can be reused in different settings.

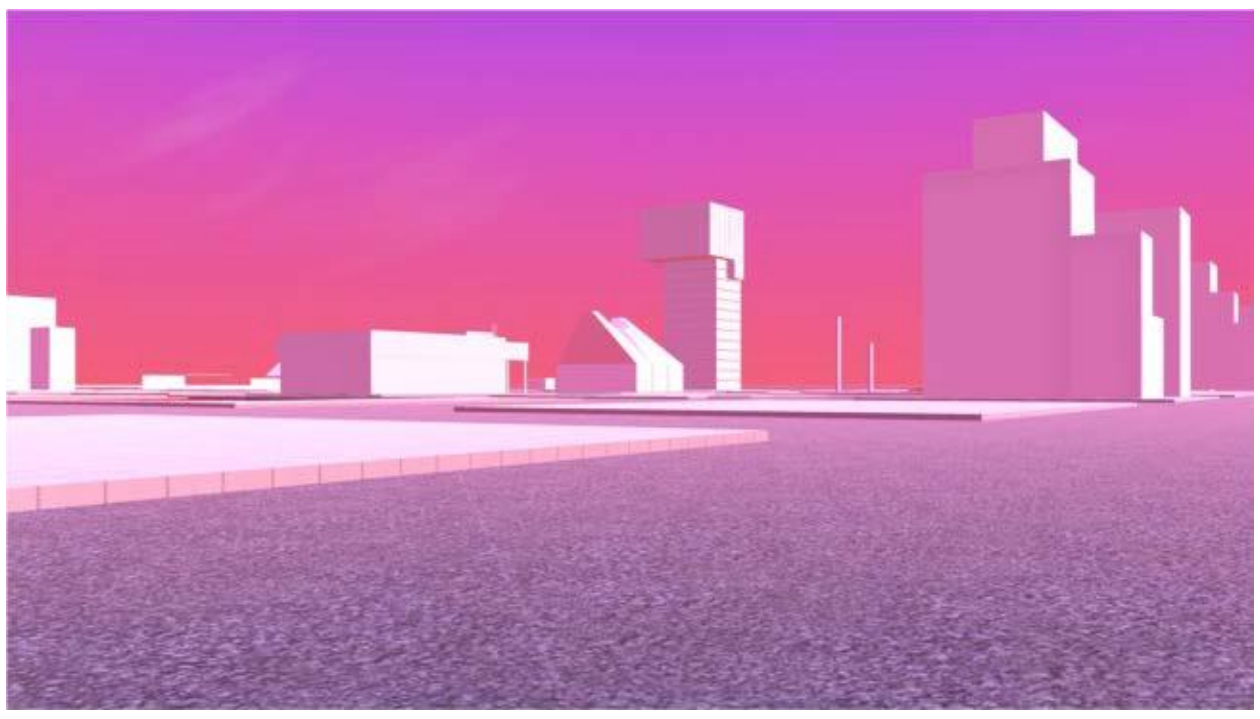

Figure 3: Main Scene

## 12 Research results

The research evaluation of the game included two phases: the pilot and the main phase. The pilot implementation was carried out on a sample of eighteen (18) students of the Department of Sociology of the University of Athens. After taking the feedback into account, it was decided to reduce the number of episodes to twenty-nine (29) and the spatial types to fourteen (14), as well as to rewrite the descriptions and the proposals of the available options-reactions. One hundred and twenty-eight (128) people participated in the main phase of the survey: students of higher education institutions, students of general high school and vocational high schools, students of technical schools. The sample consists of 50.4% men, 41.6% women, 4.8% people who identify themselves as ”other” and 3.2% people who do not wish to identify themselves in terms of gender identity. It should be noted that the interpretation and analysis of the reaction-response was carried out on the basis of an initial approximate typology that distinguished between logothetic formations that reflect political-social attitudes. In the episodes with the five possible reaction-response formations, these formations are: Democratic Radicalism (DR), Liberalism (F), Apoliticalism (A), Conservatism (C), and Authoritarianism (AUT). In the episodes with the six possible reaction-response, the formations are: Humanism, Communitarianism, Meritocracy, Technocracy, Realism, Cultural Reductionism and Authoritarianism (Autarchism). Following are some key points from the analyses of the data for issues that were in the public eye at the time of writing the paper.

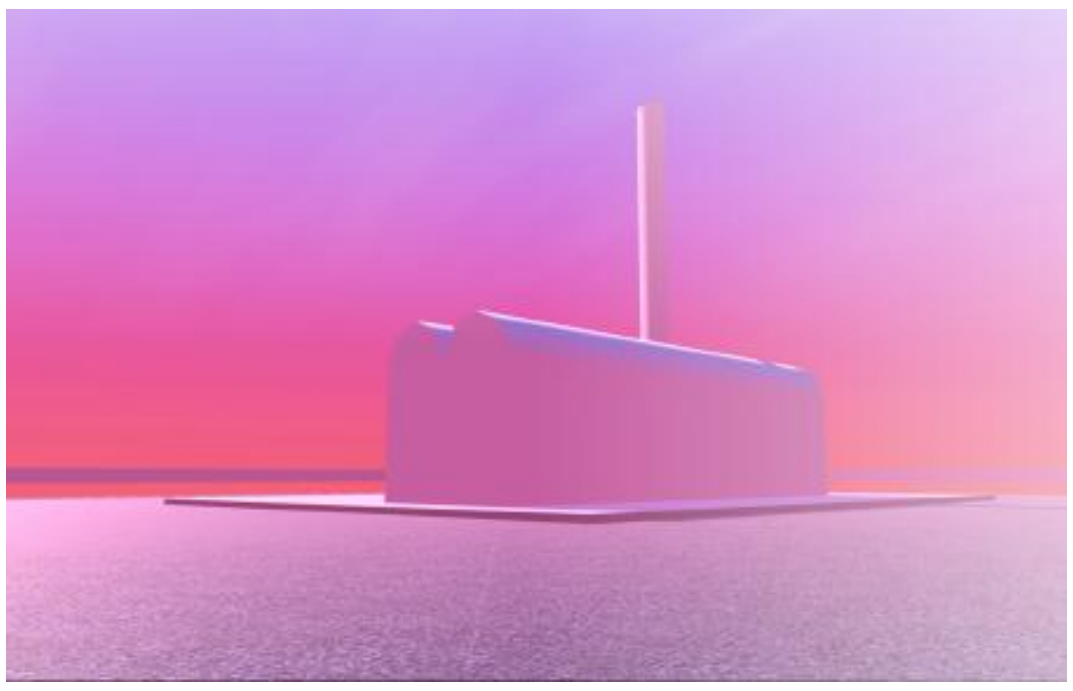

(a) A-political Factory instance.

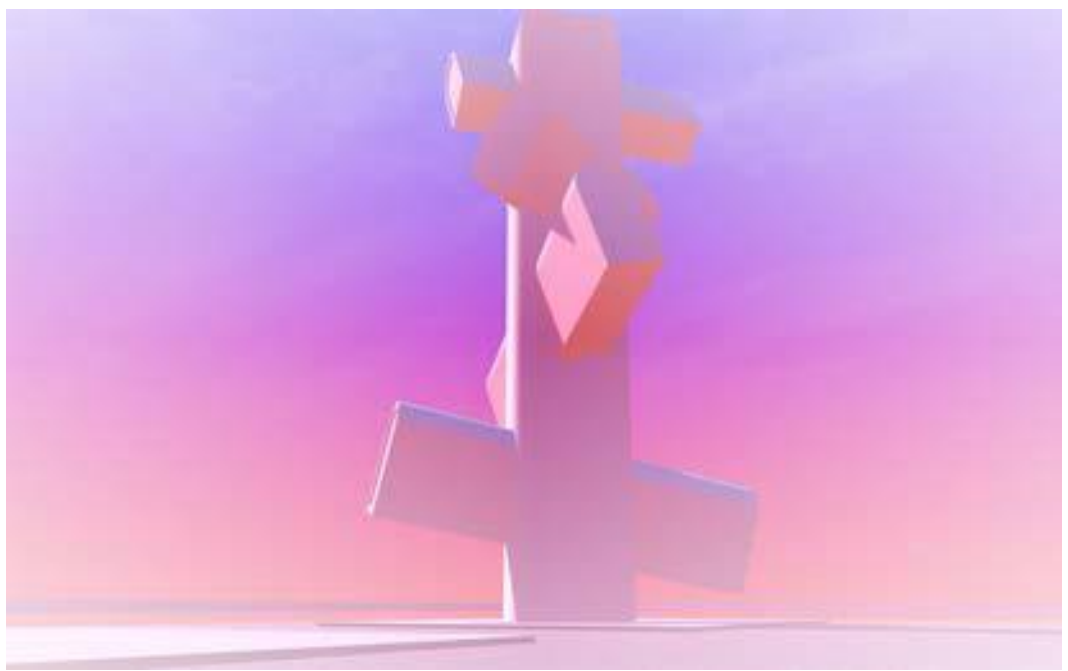

(b) Conservative Factory instance.

Figure 4: Factory instances

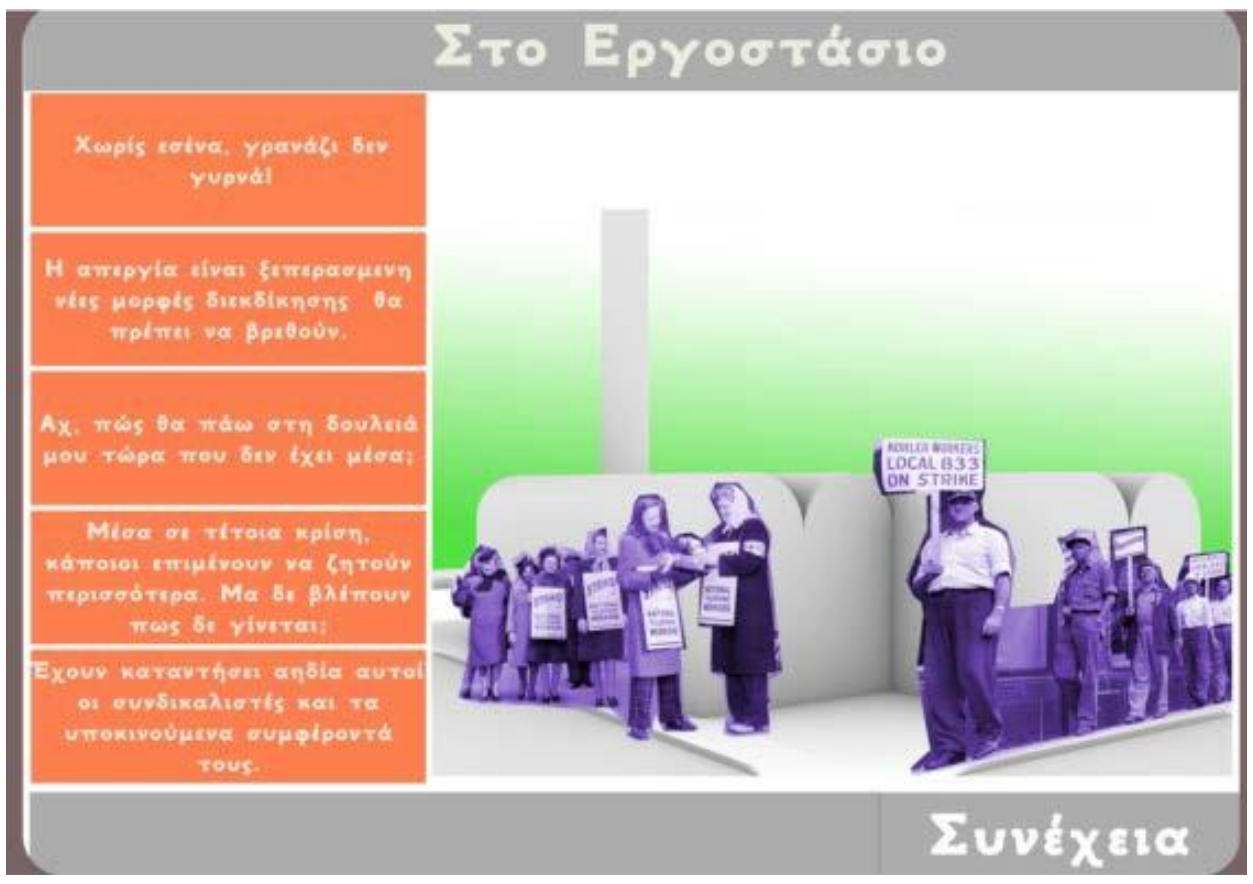

Figure 5: Factory Dilemma 1

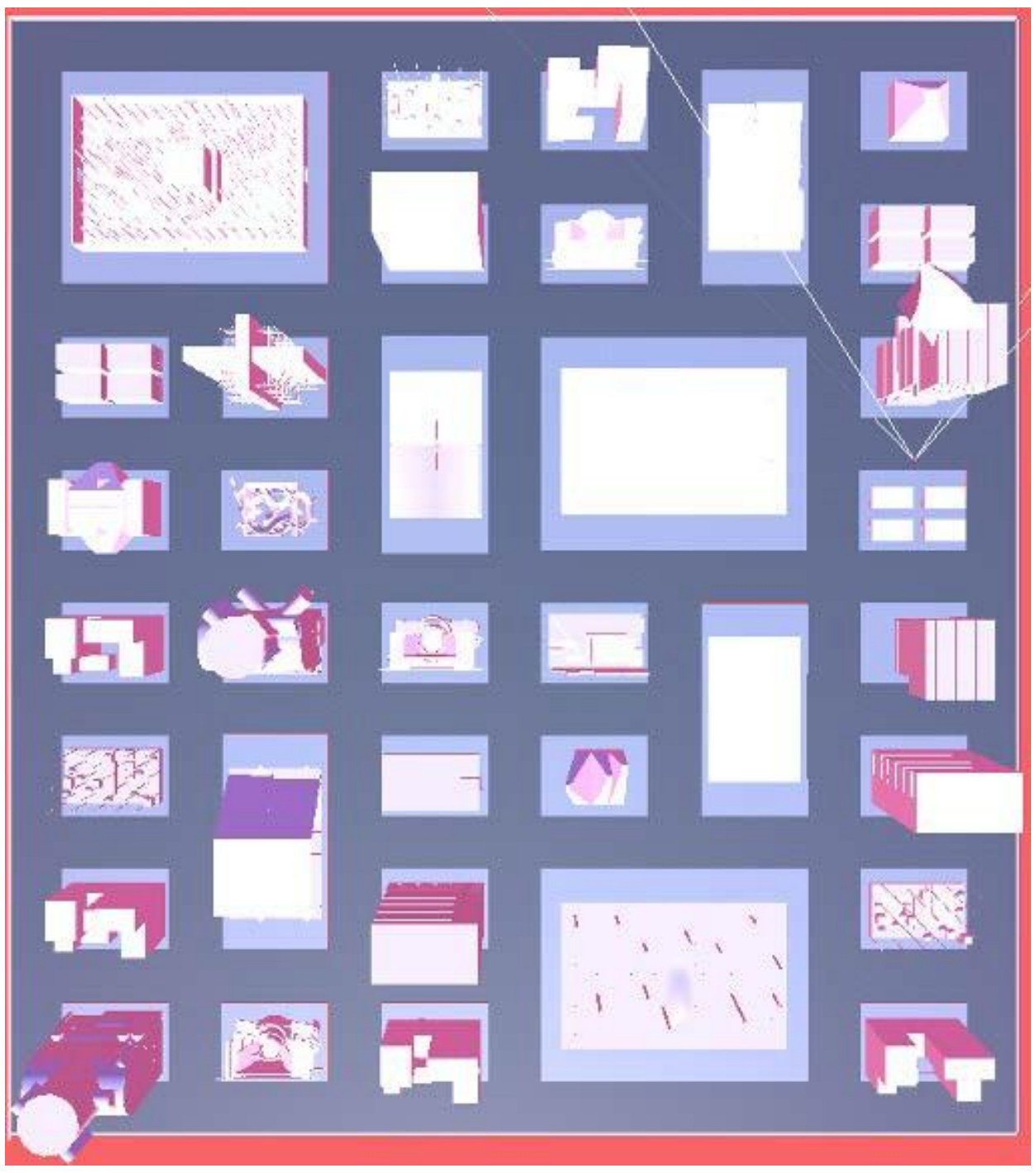

Figure 6: City-state from above

## 12.1 Prevailing reaction-responce

The vast majority of responses belong to the Democratic Radicalism and Liberalism formations. More specifically, with respect to sex and sexuality: about 49% of the sample, agree with marriage between same-sex couples, 48.4% welcomes the ”immediate reaction” of female students following a complaint of sexual harassment, 88.3%, has a positive attitude about public breastfeeding while 84.3%, exhibit progressive, feminist reflexes in relation to a disclosure of a woman’s decision to terminate a pregnancy. With respect to religion: 54% argue that it is the right of each pupil to do as he or she wishes about the morning prayer at school (F), 38.1% reacts to the priest’s unscientific sermon by demanding an effective separation of church and state(F) while 25.4% support the right of everyone to say what they want (A). 49.2% attribute the motives of an attack on a church to fanaticism, which they denounce ”wherever it comes from” (F). With respect to policing and state violence: 53.3% disagree with the presence of police in universities(PP), 38.3% show empathy and understanding in the face of pursuing a person who has committed petty theft in the market (PP), 48.8% express anger at the repeated incidents of police violence. Faced with the issue of being checked when entering the park (PP) 32.8% resent it (DP) while 37.5% want discretionary security (F). In relation to rights in public spaces: 54.7% defend the right of all to coexist in public space (PP), 41.7% prioritise the protection of open public spaces (PP), 29.9% recognise both the importance of supporting business (e.g., the extension of table seating for a private catering business in the square) and the need for children to have space to play (F),68% of the sample supports the right of owners and pets to play freely (PP). Regarding social demands, and more specifically the gathering of leftist and feminist organizations to protest in the square, 47.2% accept the right of citizens to protest (F). 47.2% accept the right of citizens to protest (F), as for the factory strike, 44% consider it obsolete, supporting the need for new forms of struggle (F), 33.6% support the squat at the university and the struggles of the students (PP) while 29.6% do not agree with the squat but understand the reasons for the protest (F). Regarding migration: when viewing the image of a street sponger in the market 63% of the sample stresses the importance of social and economic integration of marginalized groups (DP). When facing the image of a queue of people waiting to apply for asylum, 29.9% of the sample supports the ideal of harmonious and equal multicultural coexistence (communitarianism) while 28.3% focus on the fact that migrants and refugees are tortured people (humanitarianism). Similarly, 62.5% of participants feel regret about the exclusion of refugee children at school (humanism). Ragarding conflicts: in relation to youth violence, 35.2% of the sample prioritises the cultivation of empathy at school (humanism) while 34.4% pin their hopes for dealing with the phenomenon on school counsellors (technocracy). In the face of reactions outside the theatre over the ”blasphemous” theme of the show, 38.1% defend tolerance of freedom of artistic expression (F) while 27.8% oppose censorship, considering such reactions regressive (PP). Regarding economy: Faced with a company’s decision to staff the management with well-paid managers from abroad at the expense of staff salaries, 28.6% of players react by expressing solidarity with the workers (communitarianism) while 27.8% agree with the view that everyone should be paid according to their contribution (meritocracy). Regarding the eviction of small business owners, 42.1% worry about the future of people losing their jobs (humanitarianism). Regarding Environment and Culture: In the episode of environmentalists protesting outside a factory that pollutes the environment, 33.1% of participants put forward community responsibility (of each neighbourhood) (communitarianism) while 24.4% support the green transition of the economy (meritocracy). Faced with the issue of archaeological finds during excavations raise the need to stop redevelopment plans, 31.2% support the need for public consultation (PP) while 29.6% point to the multiple benefits to the country from the enhancement of cultural heritage (F).

## 12.2 Clustering

During a second experiment, player responce-reaction was first explored with histograms e.g., (Figure 7) and spider plots e.g., (Figure 8); next, data was clustered using the k-means clustering algorithm e.g., (Figure 9). First, Principal Component Analysis (PCA) was used for dimentionality reduction, i.e., in order to identify the minimum number of components required to explain the majority of the variance (97%) in the data, which returned a silhouette score of 0.4165, which corresponds to 9 clusters. K-Means, was employed next to identify patterns on the PCA-transformed data. Each data point is assigned to a cluster, providing insights into group behaviors within the data as shown in Figure 9.

To cross-check the results, the data was also processed by the Uniform Manifold Approximation and Projection (UMAP), a non-linear dimensionality reduction technique, which, compared to PCA, often provides more intuitive visualizations, particularly when dealing with highly complex datasets. K-Means, was employed next to identify patterns on the UMAP-transformed data as shown in Figure 10. To compute the number of clusters, first the dataset was reduced to three dimensions using UMAP and next both the elbow and silluette were calculated resulting in 10 clusters.

## 12.3 AI

A third experiment explored an AI solution capable of classifying data to predict the gender of each participant based on their answers. To validate our analysis, we used *lazypredict,* a Python library designed to simplify the process of model selection by training and evaluating multiple machine learning models with minimal setup. It provides a quick baseline for comparison and insights into model performance both for the initial dataset and the normalized PCA/UMAP vectors. More specifically, we defined three classes for prediction: man, woman, and other and reduced the dimentionality to three dimensions. We relabeled the data classes and adjusted their frequencies using np.unique values and then executed our AI system simulation across various models. The results and accuracy of our AI system are displayed in Table 1 achieving approximately 70% accuracy.

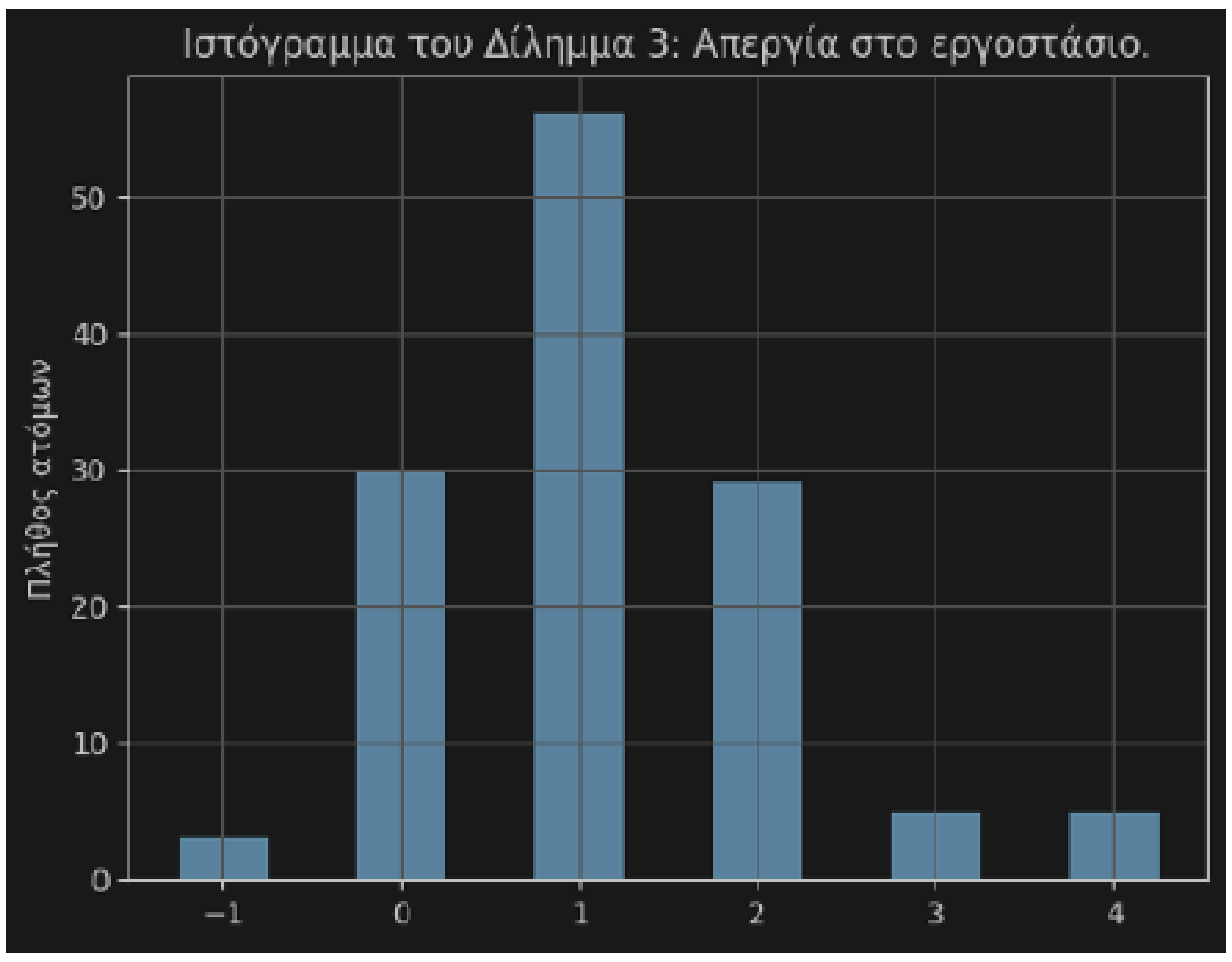

Figure 7: Factory challenge response-reaction histogram.

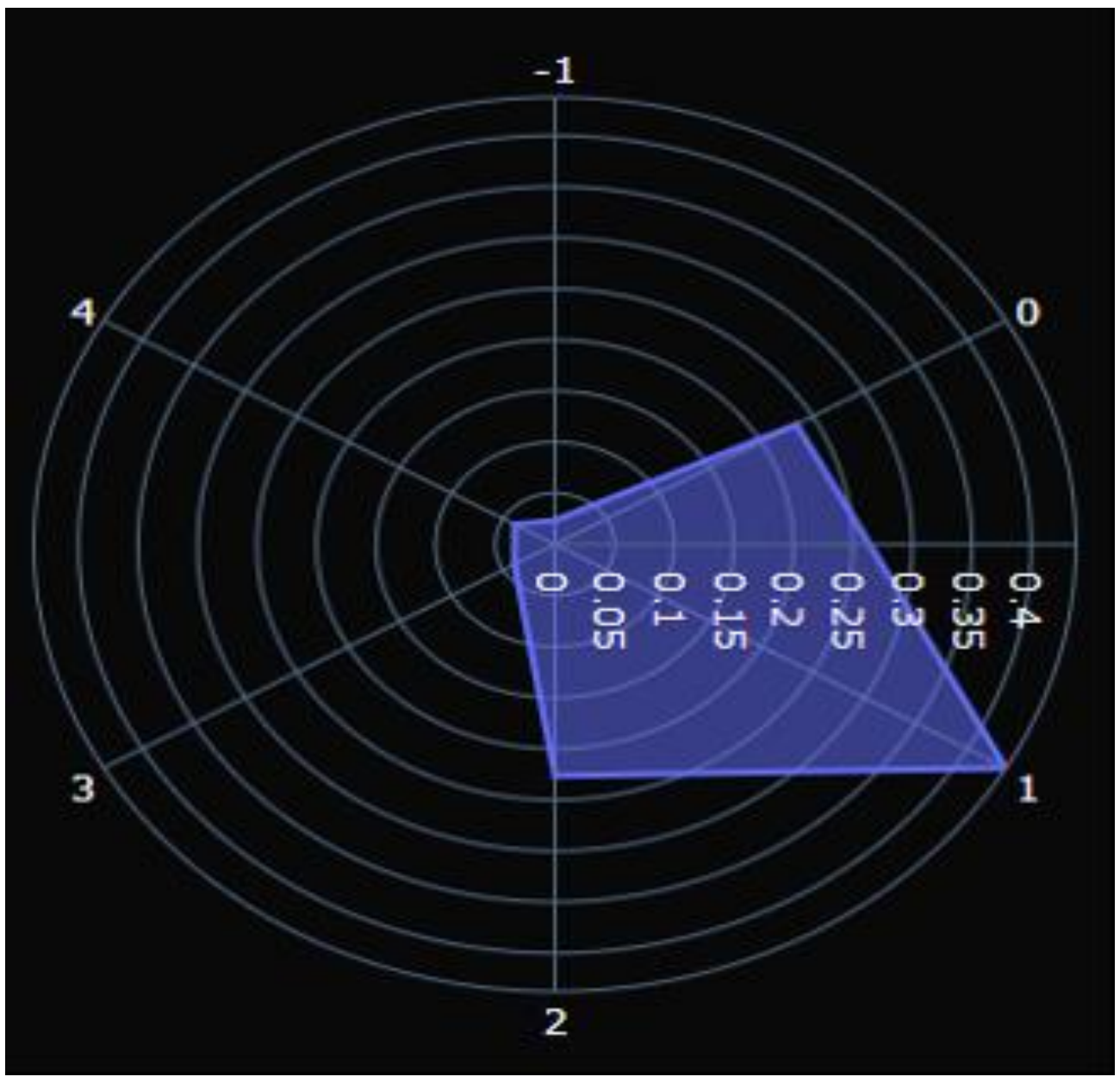

Figure 8: Factory challenge response-reaction spider plot

## 13 Towards Big Data

In order to be able to test our research hypotheses using a large number of players we created e-polis sim, a simulation model of e-polis that belongs to the category of multi-agent models. The Netlogo program was chosen for the simulation as it has a simple and flexible programming language, easy syntax and the ability to create commands from the user. The model is built through fixed elements ”patches” and mobile ”turtles” the number of which is unlimited. Both mobile and fixed elements are graphically displayed on the program desktop. The user has the ability to interact with the model directly from the control center as well as with buttons and scroll bars during iterations.

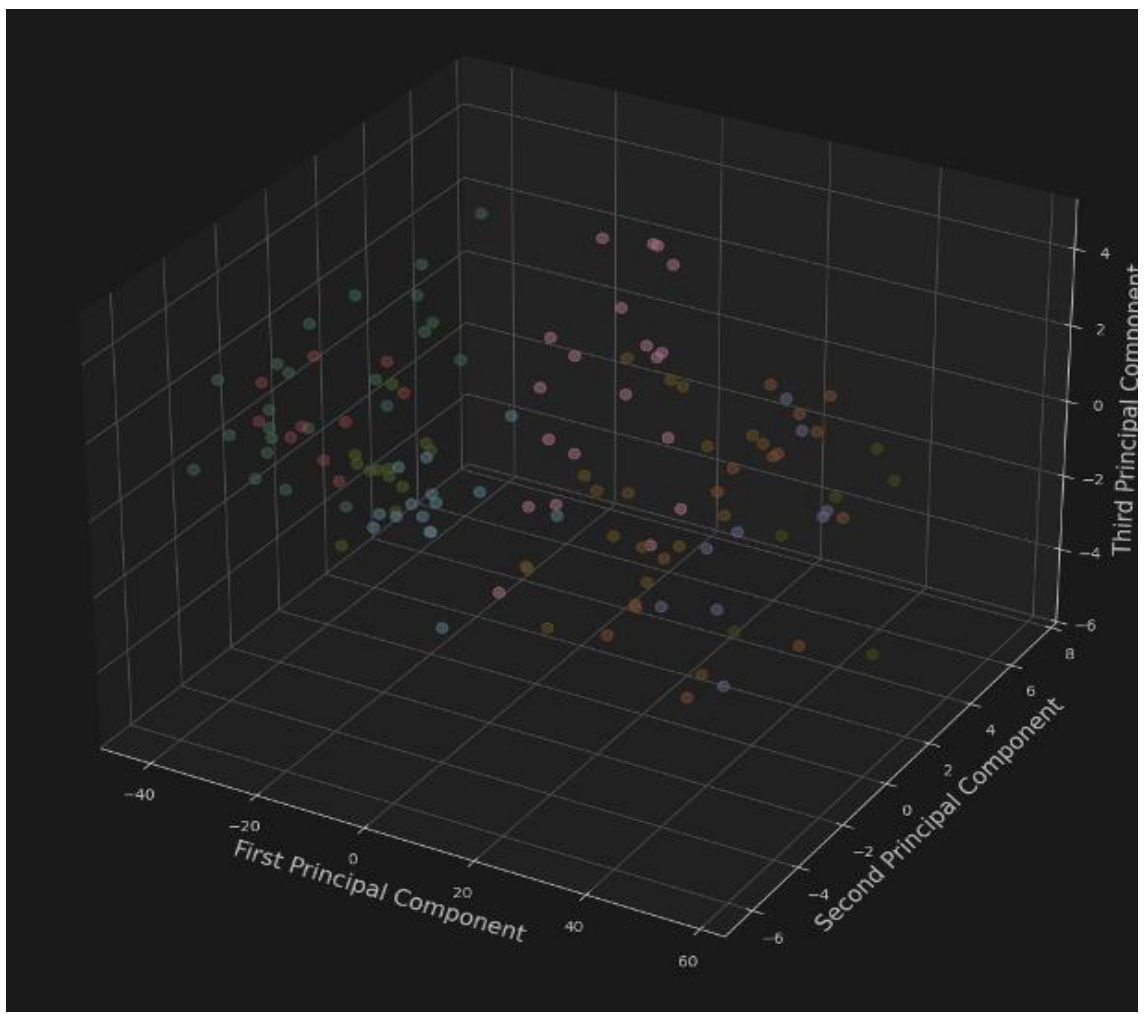

Figure 9: e-polis response-reaction k-means clusters (PCA)

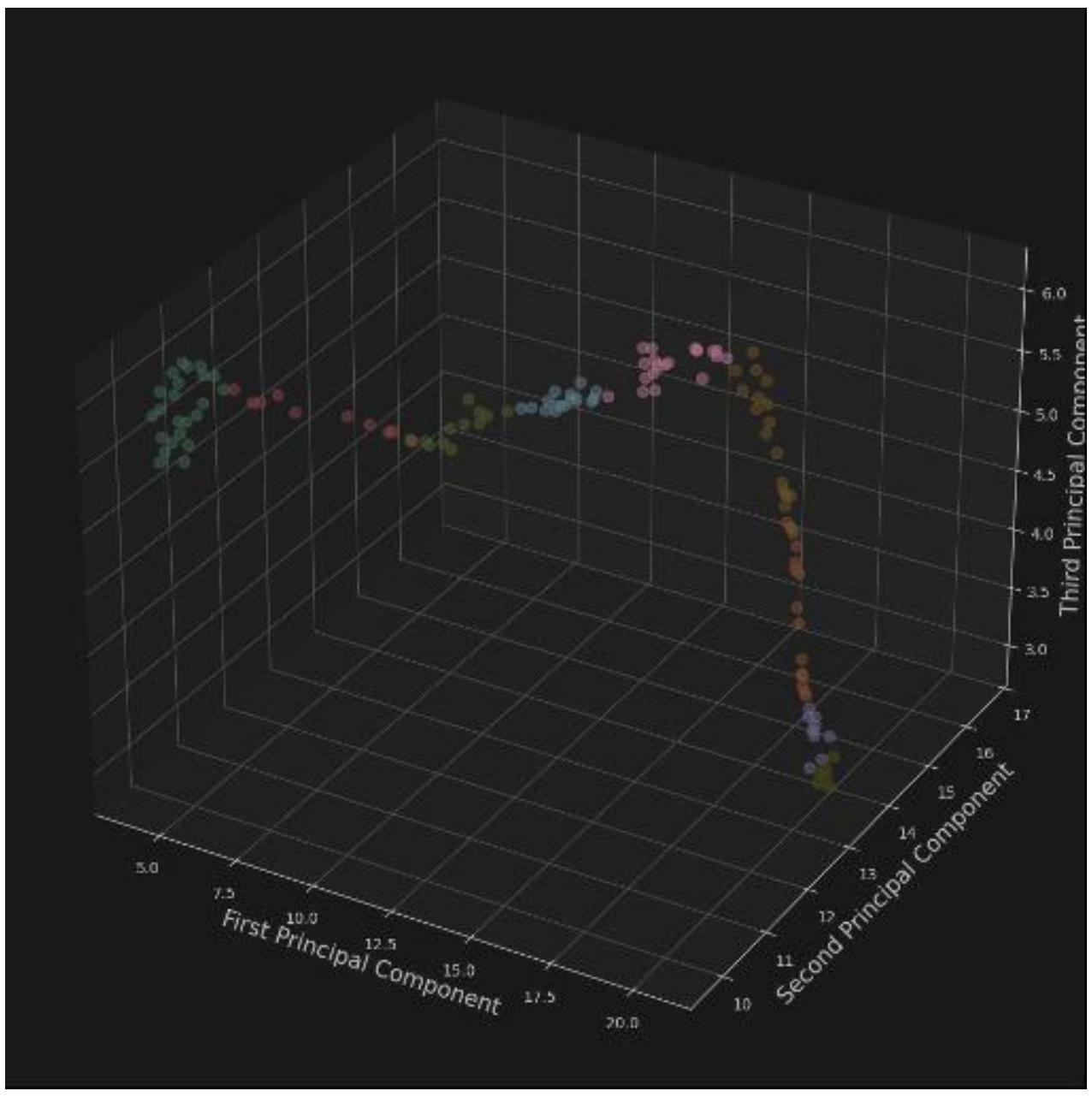

Figure 10: e-polis response-reaction k-means clusters (UMAP)

| Model | Accuracy | Balanced Accuracy | ROC AUC | F1 Score | Time Taken |
|---|---|---|---|---|---|
| RandomForestClassifier | 0.78 | 0.58 | None | 0.69 | 0.01 |
| BernoulliNB | 0.75 | 0.55 | None | 0.73 | 0.01 |
| SVC | 0.72 | 0.53 | None | 0.69 | 0.01 |
| ExtraTreesClassifier | 0.72 | 0.53 | None | 0.69 | 0.05 |
| GaussianNB | 0.72 | 0.52 | None | 0.7 | 0.01 |
| PassiveAggressiveClassifier | 0.69 | 0.51 | None | 0.66 | 0.01 |
| XGBClassifier | 0.69 | 0.51 | None | 0.67 | 0.11 |
| ExtraTreeClassifier | 0.66 | 0.49 | None | 0.64 | 0.01 |
| Perceptron | 0.66 | 0.48 | None | 0.63 | 0.01 |
| RidgeClassifierCV | 0.62 | 0.46 | None | 0.6 | 0.01 |

Table 1: Evaluation of AI model predicting the gender of e-polis participants

Agents are independent entities with different characteristics that represent the members of a community roaming in the Smart City. Agents move freely in space in random directions and interact with each other and with Smart Spatial Types. It was assumed that when an agent of a given philosophical doctrine meets a Smart Spatial Type it ”infects” it by causing it to change colour.

Breeds, i.e., classes of turtles (of different philosophical doctrine as well as classes of Smart Spatial Types (e.g., factories) were created in the simulation and represented visually using shapes. The game is parameterised with respect to the inputs:

- the number and distribution of political breeds of agents, the number and distribution of simulated institutions (Smart Spatial Types) and the number and distribution of challenges
- the pattern of agent movement that ranges from random walk to pedestrian simulation
- the ”dice”, i.e., the probability with which an agent of breed Liberalism (F) causes a Smart Spatial Type to change colour to yellow (the colour representing Liberalism in the simulation) after approaching it.

The game has the following outputs:

- the state to which the smart city converges after a sufficient number of iterations; it shows the impact of the agents philosophical belief on the Smart City
- the distribution of politicised Smart Spatial Types
- the descriptive statistics of the random function of the simulation
- the distribution of the ”happiness” function of the agents.

The desired number of instances of these classes were created programmatically at the setup phase of each experiment and they populate the ”scene” according to some initialisation scenarios. For example, in one scenario, their initial positions are at the bottom left of the scene, while in another they start moving from initial locations that are scattered in the city.

In one version of e-polis-sim, in addition to ordinary residents, there are also so-called leaders: Social activists who want to influence society so that many citizens gain influence in shaping the urban fabric through mutual interactions. Positive leaders are those with extreme and positive trust in people, they have a beneficial influence on social trust in others. Negative leaders show extreme and negative trust in people, thus reducing the trust of the people with whom they interact. Furthermore, there is the assumption that the mere presence of a citizen in a given neighbourhood or building affects their trust in other people. Being in an office, healthcare facility, workplace, school, or industrial or office area reduces trust (temporarily or permanently). On the other hand, being in a park, museum, highways, or in a historical or recreational area increases trust (temporarily or permanently). These changes are called location influence. The simulations carried out aim to determine how the long-term change in people’s trust (mutual and towards institutions) over time affects the level of social participation and, indirectly, the development of open civil society and deliberative democracy.

## 14 Conclusions and future work

Serious game development requires a multitude of skills both in the fields of creativity and IT; in one word, an interdisciplinary approach. At the front end and, in depth knowledge of concept modelling, UI design, animation, 3D modelling; at the back end, event-based programming, AI, databases, cloud computing, web services and desktop and mobile computing workflows are essential elements required to build serious games. Furthermore, although tools such as Unity and Unreal, [75], platforms are available, they require a steep learning process while end products take up substantial computing resources such as RAM and CPU acceleration, requiring updated operating systems and advanced security. Lack of standardisation in this area forces researchers to develop games from scratch instead of reusing existing components.

This work tackles the issue of scalable, system-level, computationally-efficient abstract modeling of the *simulated physical world* in the context of the e-polis serious game. Its contributions are a formal definition of a knowledge representation schema for Smart City-States as well as a mechanism for reasoning with such knowledge using logical deduction that combines expressiveness, scalability and performance, by leveraging the dynamic knowledge-base maintenance system introduced in [73]. The deductive layer of this system stores abstract knowledge on players' preferred political position in specific socio-political dilemmas, as well as on the level of interest of players. for smart spatial types. Furthermore, historical predicates of such knowledge mined with AI techniques can generate knowledge on the overall political position of a user, as well as a characterisation of the most prevailing political flavour of significant Smart Spatial Types. In this view, player location, duration of visit, political choice act as digital *biomarkers* for both players and Spatial Types.

Furthermore, we have created the e-polis Game that comprises the following game assets:

- a library of 3D Smart Spatial Types, that represent selected institutions, namely: *Agora (Market), Archaeological site, Cinema, Factory, Office building, Government building, Police station, Park, School, Theatre, Temple, University, Urban void*.

- Five political categories of content corresponding to the philosophical doctrines: *Apoliticism, Democratic Realism, Critical Radicalism, Conservatism and Autocracy*.

- a set of Dilemmas, namely: *travelling salesman, small businesses, theft, dispute, play, strike, ecology, gay wedding, leave to remain, immigrants, pets, surveillance, hook-up, violence, prayer, bullying, censorship, demonstration, abortion, preaching, squatting, asylum, abuse of power*.

- a customisable blueprint of a smart city-state.

- a customisable blueprint of a socio-political dilemma.

- an open-source implementation for the Unity Game. Engine.

In terms of future work, we are working towards integrating more AI analytics over the collected data in order to derive richer knowledge such as the clustering of the users to opinion groups. We have also developed a agent-based approach of e-polis which is the subject of a forthcoming publication. An interesting research direction is also the integration of machine-readable semantically annotated politicised material, both 3D Spatial Types and statements. Such material is available through research teams that work on IR techniques such as Sentiment Analysis and used to keep the dilemmas up to date with most current affairs, [76]. Although this approach is focused in the requirements of the e-polis Game, it can be used as a base for serious game development in the context of Smart City States in general.

*Acknowledgments*

This work was funded in part by the research project "e-polis of the future", supported by the Hellenic Foundation of Research and Innovation (H.F.R.I.) in the context of the "1st Call for H.F.R.I. (http://www.elidek.gr) Research Projects to Support Faculty Members & Researchers and Procure High-Value Research Equipment" (Project Number: 2617).

# Appendix

## A Unity basic components

A prototype implementation of the specified system was developed using the Unity game engine. *Scenes* are assets that contain all or part of a game or application. *Cameras* are used to display the game world to the player. In Unity, you create a camera by adding a Camera component to a GameObject. *GameObjects* are the fundamental objects in Unity that represent characters, props and scenery. They act as containers for *Components*, which implement the functionality. Unity has lots of different built-in component types, while custom components can be developed using the *Unity Scripting API*. Scripts allow for triggering game events, modifying Component properties over time and responding to user input. Unity supports the C# programming language natively. The *Transform* component determines the Position, Rotation, and Scale of each object in the scene. Every GameObject has a Transform. A *Prefab* Asset acts as a template from which new Prefab instances can be created in the Scene. It is a resusable asset that stores a GameObject complete with all its components, property values, and child GameObjects. An *Asset* is any item used in a Unity project to create a game or app. Assets can represent visual or audio elements, abstract items such as color gradients, animation masks, or arbitrary text or numeric data for any use. Some types of asset can be created in the Unity Editor, such as a ProBuilder Mesh, an Animator Controller, an Audio Mixer, or a Render Texture.

### A.1 Unity UI

The *Unity User Interface*is a GameObject-based UI system that uses Components and the Game View to arrange, position, and style user interfaces. The *Canvas* is the area that all UI elements should be inside. The Canvas is a Game Object with a Canvas component on it, and all UI elements must be children of such a Canvas. Every UI element is represented as a rectangle for layout purposes. *Visual Components* help you create GUI specific functionality. The *Text* component, which is also known as a Label, has a Text area for entering the text that will be displayed. It is possible to set the font, font style, font size and whether or not the text has rich text capability. An Image has a Rect Transform component and an Image component. *Interaction components* handle interaction, such as mouse or touch events and interaction using a keyboard or controller. They are not visible on their own, and must be combined with one or more visual components in order to work correctly. A *Button* has an *OnClick* UnityEvent to define what it will do when clicked. A *Toggl*e has an *Is On checkbox* that determines whether the Toggle is currently on or off. A *Slider* has a decimal number Value that the user can drag between a minimum and maximum value. A *Scrollbar* has a decimal number Value between 0 and 1. When the user drags the scrollbar, the value changes accordingly. A *Dropdown* has a list of options to choose from. All three also have a *OnValueChanged* UnityEvent to define what it will do when the value is changed. An *Input Field* is used to make the text of a Text Element editable by the user. It has a UnityEvent to define what it will do when the text content is changed, and an another to define what it will do when the user has finished editing it.

### A.2 Unity Event System

The *Event System* is a way of sending events to objects in the application based on input, be it keyboard, mouse, touch, or custom input. The *Event System* consists of a few components that work together to send events and its primary roles are managing which GameObject is considered selected, which Input Module is in use and updating all Input Modules as required.

### A.3 Unity Physics Engine

In Unity, a *collision* happens when two GameObjects that are configured for collision occupy the same physical space. Collision is a foundational part of most games, and many interactive applications and simulators. To handle collision between GameObjects, Unity uses *colliders*. A collider is a Unity component that defines the shape of a GameObject for the purposes of physical collisions. Colliders are invisible, and do not need to be the same shape as the GameObject's mesh. The *OnTriggerEnter()* function is a widely used function in the Unity game engine [68]. It is used to detect when an object enters a trigger collider attached to a GameObject in a Unity scene. The *OnTriggerEnter()* function is part of Unity's scripting system, which allows game developers to add interactivity to their games by writing code in various programming languages. When a GameObject with a trigger collider enters the area defined by the trigger collider, the *OnTriggerEnter()* function is called. This function is typically used to initiate some kind of action or behavior in response to the trigger event. For example, a trigger collider could be used to detect when a player enters a particular area of a game level, triggering an event such as a cutscene, the appearance of enemies, or the revelation of a hidden item. In order to use the *OnTriggerEnter()* function in Unity, a script must be written that associates the GameObject object with the trigger configurator. The script must include a method called

*OnTriggerEnter()*, which takes a unique parameter of type Collider. This parameter represents the collider of the object that has entered the trigger zo Overall, the *OnTriggerEnter()* function is Overall, the *OnTriggerEnter()*function is a powerful tool in the Unity game engine that allows developers to create dynamic and interactive game experiences. Using trigger colliders and the *OnTriggerEnter()* function, game developers can create complex game mechanics and events that respond to player actions in real-time. Unity also provides the *OnTriggerExit()*, *OnTriggerStay()* related functions.

### A.4 Scripting concepts

Running a Unity script executes a number of event functions in a predetermined order. Figure 11 describes those event functions and shows how they fit into the execution sequence. A script makes its connection with the internal workings of Unity by implementing a class which derives from the built-in class called *MonoBehaviour*.The *Update()* function is the place to put code that will handle the frame update for the GameObject. To enable the Update() function to do its work, it is often useful to be able to set up variables, read preferences and make connections with other GameObjects before any game action takes place. The *Start()* function will be called by Unity before gameplay begins (ie, before the Update function is called for the first time) and is an ideal place to do any initialization. The construction of objects is handled by the editor and does not take place at the start of gameplay as one might expect and therefore constructors cannot be defined for a script component.

## B e-polis front-end

The game consists of three Scenes:

- the *Welcome scene*, that contains basic instructions, a Play Button and a Quit Button
- the *Main scene* that hosts the Smart City-State
- the *Dilemma scene* that hosts all Dilemmas.

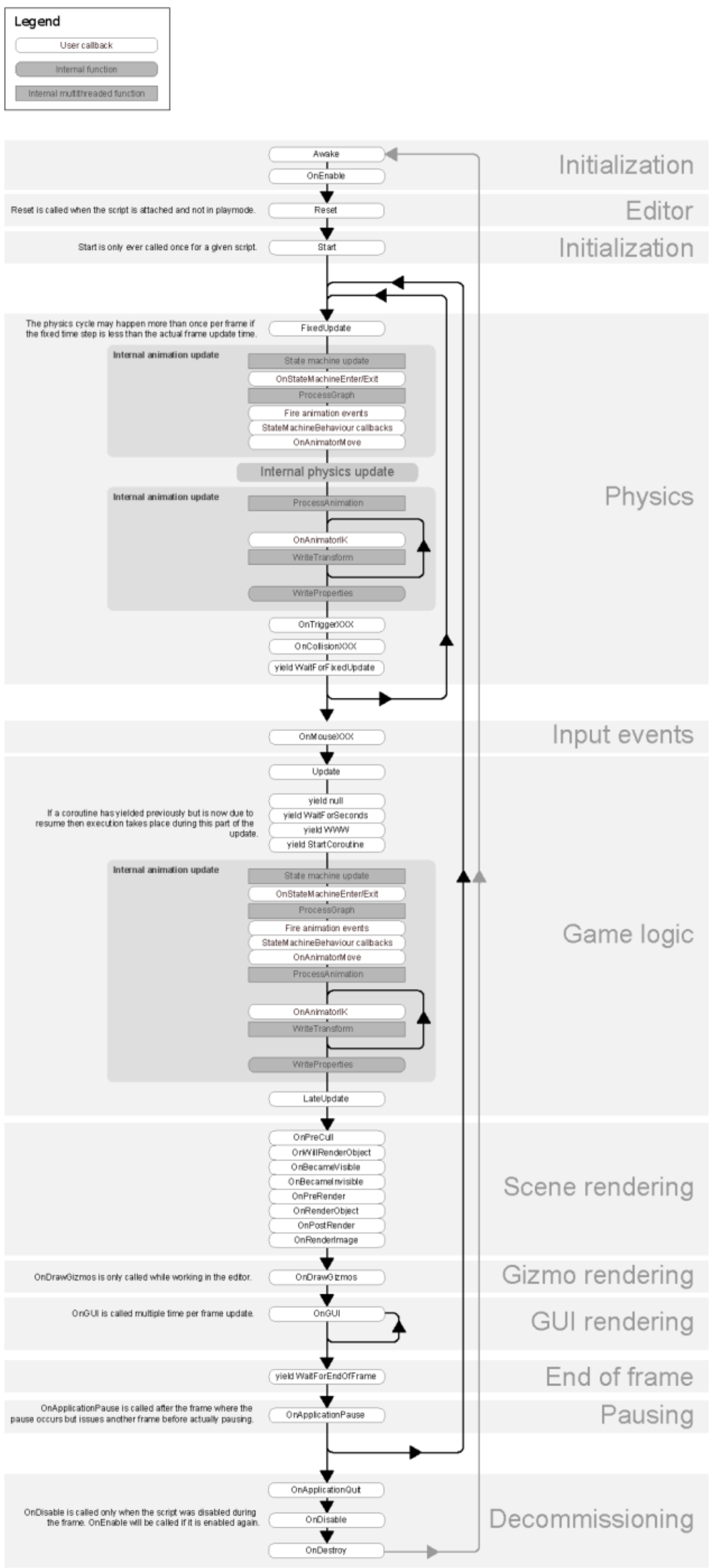

Figure 11: Monobehaviour lifecycle

### B.1 Main Scene

For the purpose of constructing the Smart City-State layout that is part of the Main Scence, SpatialTypes, RoadBlocks and Dilemmas are the main building blocks and therefore they were constructed as Prefabs. *SmartSpatialTypes* comprise of *Asphalts* and *Sidewalks* as well as five *PoliticisedSpatialTypes* that are 3D objects representing the building infrastructure of the *SmartSpatialTypes*. The latter were developed in the Rhino software and were exported in *object* format in order to be imported into Unity where they appeared as Prefabs. Because each Prefab was a complex object that had a hierarchical structure consisting of a title, spatial type, material and as a result it was not recognized by Unity, they were deconstructed into a flat hierarchy and only the 3D objects were retained, as GameObjects with a Transform. *Asphalts* and *Sidewalks* and *Roadblocks* are also PolyhedralSpatialTypes, i.e., gameobjects with a Transform that represents their location and size. *Neighborhouds* are Prefabs that contain six *SmartSpatialTypes* with roads in between. These prefabs were copied to create a Smart City State with thirty-six (36) building blocks, i.e., *SmartSpatialTypes*.

Next, in order to add interaction, each *PoliticisedSpatialTypes* was added two BoxCollider components, one nested in the other. The Box collider is a built-in cube-shaped collider. The *Is Trigger* property was enabled to use the collider as a trigger for events. When Is Trigger is enabled, other colliders pass through this collider, and trigger the messages *OnTriggerEnter*, *OnTriggerStay*, and *OnTriggerExit*.

### B.2 Dilemma Scene

Each Dilemma is a Prefab that contains a Canvas with a back, a top and a bottom Panel, an Image, Five Buttons for each availabe choice and one Button for Continuing the Game. The Canvas has a transform that stored its position. When the Scene is loaded, the Main Camera Transfer is passed as a parameter so that the user has a full-frontal view of the Dilemma and can interact with it.

## C e-polis back-end

The e-polis programming logic is based on C# scripts that belong to three categories depending on their scope.

- Game scope: *QuitManager.cs*, *RetrievePlayerPosition2.cs*, *RetrieveCameraPosition2.cs*, *< SmartSpatialType >DilemmaManager.cs*,
- SmartSpatialType scope: *< SmartSpatialType > InstanceManager.cs*.
- Dilemma scope: *DilemmaCanvasManager.cs*

Game scope is required for a) Quit() and b)ViewFromAbove and in general for functionality specified in *Rules 2c, 3b, 3c, 3d, 3e, 4* (Section D) where scripts from one scene have to coordinate with scripts of another. Game scope scripts are attached an inactive GameObject e.g., a SideWalk GameObject in the Main Scene. As the default scope for all scripts attached to GameObjects is GameObject scope, in order to implement Game scope, it was necessary to define global variables stored in the PlayPreferences data structure. These global variables implement the e-polis predicates: *L_UserChoice()* as well as global constant predicates such as predefined *L_AtomicLocation()* both for SmartSpatialTypes and Dilemmas. For example *SmartSpatialType > DilemmaManager.cs* coordinates the state of the five collocated *PoliticisedSpatialTypes* so that only the one that corresponds to the most recent user choice is enabled.

SmartSpatialType scope is required for functionality specified in *Rules 1a, 2a, 2b* (Section D). Scripts in this category are attached to *PoliticisedSpatialTypes*.

Scripts in this category determine what happens whenever the player enters each of the active zones (both when the spatial type is enabled or disabled) by calling *OnTriggerEnter(), OnTriggerExit()*. It also determines what happens when the dilemma is triggered and when the dilemma has been completed and player returns home.

Dilemma scope include the functionality specified in *Rule 3b* and it is implemented by the *DilemmaCanvasManager.cs* that is attached to the Canvas GameObject of each Dilemma. Last, but not least, the player's view is through the eyes rather than from above, and was implemented with the First Person Controller (FPC) library of the Unity StarterAssets package.

As such, regarding **e-polis Monobehaviour**, the scripts here invoke the *Start()* function every time the Main Scene is loaded. This happens in the beginning of the game as well as every time the user responds to a Dilemma.

### C.1 Main Scene: Start()

The tasks that take place are:

- the coordination of the object references of all five Politicised instances of the same Smart Spatial type and setting their state so that only the one that matches the latest user choice of the linked dilemma is enabled and the rest are disabled.
- the retrieval and setting of the user location/rotation (Transform) from the UserPrefereces so that it looks at the Smart Spatial Type transformation. When the user has asnswered all Dilemmas, his/her location/rotation is set to a ViewFromAbove predefined location to look at the city from above.

### C.2 Main Scene: OnTriggerEnter()

The tasks that take place are:

- Detection of the player location with respect to the two active zones.
- If located in the outer zone and the spatial type instance is invisible, the latter becomes visible.
- If located in the inner zone, a dilemma is triggered

### C.3 Main Scene: OnTriggerExit()

The tasks that take place are:

- Detection of the player location outside the smart spatial type. The visibility of the spatial type instance is not affected.

### C.4 Main Scene: $Load < SpatialTypeDilemma > ()$

The tasks that take place are:

- setting player's return position and rotation (PlayerPrefs)
- setting the dilemma camera next position (PlayerPrefs)
- SceneManager.LoadScene("Dilemma_scene");

Dilemma Scene scripts invoke the *Start()* function every time the Dilemma Scene is loaded. This happens when a Dilemma is triggered by the user entering the inner zone of a smart spatial type.

### C.5 Dilemma scene: Start()

The tasks that take place are:

- the retrieval and setting of the Main Camera location/rotation (Transform) from the UserPreferences to portray the Canvas of the current Dilemma.

### C.6 Dilemma scene: OnButtonClick()

The tasks that take place are:

- the assertion of a UserChoice predicate in the PlayerPreferences.
- the logging of the UserChoice predicate in Firebase.
- SceneManager.LoadScene("Main_scene"

With respect to high-level logical predicates, their most recent instance is made persistent in the User Preferences data structure. With respect to historical predicates, a time-series of instances are stored in the Firebase cloud Database. Proximity interaction is a collision between two rigid body game objects, i.e., the player avatar and the Politicised Spatial Type. To achieve this, two box colliders are wrapped around each PoliticisedSpatialType.

## D Rules

**Rule 1a: User entered the *SmartSpatialType* outer zone, the first "active" zone. If this is the first time the user approaches, the default instance has been enabled inside a previously empty building block.**

$$(L_UserAtLocation(uid\,?uid)(x\,?x)(y\,?y)(z\,?z) \wedge (L_SpatialType(sid\,?sid)(type?type)) \wedge (L_AtomicLocation(rid\,?rid)(sid\,?sid)) \wedge (L_PoliticisedSpatialType(siid\,?siid)(sid\,?sid)(pid\,?pid)) \wedge IsDisabled(siid\,?sid)(sid\,?sid) \wedge (L_UserChoice(cid\,?cid)(did\,?did)(sid\,?sid) \Longrightarrow (H_UserInLocation(uid?uid)(sid?sid)(st?st))$$

**Rule 1b: User interacts with the existing *PoliticisedInstance* of the *SmartSpatialType*.**

$$(H_UserInLocation(uid\,?uid)(sid\,?sid)(st\,?st)) \Longrightarrow (H_UserAtSmartSpatialType(uid\,?uid)(sid\,?sid)(st\,?st))$$

**Rule 2a: User entered the SmartSpatialType outer zone**

$$(H_UserInLocation(uid\,?uid)(rid\,?rid) \wedge (L_InRegion(x\,?x)(y\,?y)(z\,?z)(rid\,?rid)) \wedge (L_InRegion(nrid\,?nrid)(rid\,?rid)) \wedge (L_NestedLocation(nrid\,?nested_rid)(prid?rid)(sid\,?sid)) \wedge (L_PoliticisedSpatialType(siid\,?siid)(sid\,?sid)(pid\,?pid)) \wedge (L_SpatialType(sid\,?sid)(type\,?type)) \wedge (L_Dilemma(did\,?did)(sid\,?sid)) \Longrightarrow (H_UserInNestedLocation(uid\,?uid)(rid\,?nested_rid)(prid\,?rid)(sid\,?sid)(st\,?st))$$

**Rule 2b: User has triggered the *Dilemma***

$$(H_UserInNestedLocation(uid\,?uid)(rid\,?nested_rid)(prid\,?rid)(sid\,?sid))(st\,?st)) \Longrightarrow (H_UserInDilema(uid?uid)(did?did)(st?st))$$

**Rule 2c: User is transported to the *Dilemma* location**

$$H_UserInDilema(uid\ ?uid)(did\ ?did)(st\ ?st)) \Longrightarrow (H_UserInLocation(uid?uid)(drid\ ?dilemma_region_id)(st\ ?st))$$

**Rule 3a: Dilemma was answered by user**

$$(H_UserInLocation(uid\ \ ?uid)(drid\ ?dilemma_region_id)(st\ ?st)) \wedge (L_SpatialType(sid\ ?sid))(type?type) \wedge (L_Dilemma(did\ ?did)(sid\ ?sid)) \wedge (L_UserChoice(cid\ ?cid)(uid\ ?uid)(did\ ?did)(sid\ ?sid) \Longrightarrow (H_UserAnsweredDilema(uid\ ?uid)(did?did)(sid?sid)(st?st))$$

**Rule 3b: User is transported back to the SmartSpatialType to see the resulting transformation**

$$(H_UserAnsweredDilemma(uid\ ?uid)(did?did)(sid?sid)(st?st)) \wedge \exists(did) | \neg(H_UserAnsweredDilemma(uid\ \ ?uid)(did\ ?did)(st?st)) \Longrightarrow (H_UserInLocation(uid?uid)(rid\ ?rid)(sid?sid)(st\ ?st))$$

**Rule 3c: The *SmartSpatialType* is transformed.**

$$(H_UserAnsweredDilemma(uid\ ?uid)(did?did)(sid?sid)(st?st)) \wedge \exists(did) | \neg(H_UserAnsweredDilemma(uid\ \ ?uid)(did?did)(st?st)) \wedge (L_UserChoice(cid\ ?cid)(uid\ ?uid)(did\ ?did)(sid\ ?sid)) \wedge (L_PoliticisedSpatialType(siid\ ?siid)(sid\ ?sid)(pid\ ?pid)) \wedge (L_PoliticisedUserChoice(cid\ ?cid)(pid\ ?pid) \Longrightarrow (H_SmartSpatialTypeTransformed(tid\ ?tid)(sid\ ?sid)(siid\ ?siid)(pid\ ?pid)(st?st))$$

**Rule 3d: The PoliticisedSmartSpatialType instance that corresponds to the UserChoice is enabled.**

$$(H_SmartSpatialTypeTransformed(tid\ ?tid)(sid\ ?sid)(siid\ ?siid)(pid\ ?pid)(st\ ?st)) \Longrightarrow IsEnabled(siid?siid)(sid?sid)(pid?pid)(st?st))$$

**Rule 3e: The rest n-1 co-located PoliticisedSmartSpatialType instances are disabled.**

$$(H_SmartSpatialTypeTransformed(tid\ ?tid)(sid\ ?sid)(siid\ ?siid)(pid\ ?pid)(st\ ?st)) \Longrightarrow \forall(siid) | \neg(L_PoliticisedUserChoice(cid\ \ ?cid)(pid\ ?pid)\ IsDisabled(siid\ ?siid)(sid\ ?sid)(pid\ ?pid)(st?st))$$

**Rule 4: When all dilemmas have been answered, the user is transported to a PolyhedralSpatiaType that represents a hovering balcony with a view above the city to check the final landscape.**

$$\forall(did) | (H_UserAnsweredDilemma(uid\ ?uid)(did\ ?did)(st\ ?st) \Longrightarrow (H_UserinLocation(uid\ ?uid)(\text{"}Sky\text{"})(st\ ?st))$$

**Contribution of Individual Authors to the Creation of a Scientific Article (Ghostwriting Policy)**

The authors equally contributed in the present research, at all stages from the formulation of the problem to the final findings and solution.

**Sources of Funding for Research Presented in a Scientific Article or Scientific Article Itself**

No funding was received for conducting this study.

**Conflicts of Interest**

The authors have no conflicts of interest to declare that are relevant to the content of this article.